\documentclass[aps,pra,showpacs,floatfix]{revtex4}
\usepackage{graphicx}

\begin{document}

\title{Structure of propagators for quantum nondemolition 
systems} 

\author{Subhashish Banerjee}
\email{subhashishb@rri.res.in}
\affiliation{Raman Research Institute, Bangalore - 560 080, 
India} 

\author{R. Ghosh}
\email{rghosh@mail.jnu.ac.in}
\affiliation{School of Physical Sciences, Jawaharlal Nehru 
University, New Delhi - 110 067, India} 


\begin{abstract}
In the scheme of a quantum nondemolition (QND) measurement, an 
observable is measured without perturbing its evolution. In the 
context of studies of decoherence in quantum computing, we 
examine the `open' quantum system of a two-level atom, or 
equivalently, a spin-1/2 system, in interaction with quantum 
reservoirs of either oscillators or spins, under the QND 
condition of the Hamiltonian of the system commuting with the 
system-reservoir interaction. The propagators for these QND 
Hamiltonians are shown to be connected to the squeezing and 
rotation operators for the two baths, respectively. Squeezing 
and rotation being both phase space area-preserving canonical 
transformations, this brings out an interesting analogy between 
the energy-preserving QND Hamiltonians and the homogeneous 
linear canonical transformations. 
\end{abstract} 


\maketitle

In the scheme of a quantum nondemolition (QND) measurement, an 
observable is measured without perturbing its free motion. Such 
a scheme was originally introduced in the context of the 
detection of gravitational waves \cite{caves80}. It was to 
counter the quantum mechanical unpredictability that in general 
would disturb the system being measured. The dynamical 
evolution of a system immediately following a measurement 
limits the class of observables that may be measured repeatedly 
with arbitrary precision, with the influence of the measurement 
apparatus on the system being confined strictly to the 
conjugate observables. Observables having this feature are 
called QND or back-action evasion observables \cite{bo96, vo98, 
zu84}. In addition to its relevance in ultrasensitive 
measurements, a QND scheme provides a way to prepare quantum 
mechanical states which may otherwise be difficult to create, 
such as Fock states with a specific number of particles. One of 
the original proposals for a quantum optical QND scheme was 
that involving the Kerr medium \cite{walls}, which changes its 
refractive index as a function of the number of photons in the 
`signal' pump laser. The advent of experimental methods for 
producing Bose-Einstein condensation (BEC) enables us to make 
progress in the matter-wave analogue of the optical QND 
experiments. In the context of research into BEC, QND schemes 
with atoms are particularly valuable, for instance, in 
engineering entangled states or Schr\"{o}dinger's cat states. A 
state preparation with BEC has recently been performed in the 
form of squeezed state creation in an optical lattice 
\cite{science01}. In a different context, it has been shown 
that the accuracy of atomic interferometry can be improved by 
using QND measurements of the atomic populations at the inputs 
to the interferometer \cite{kbm98}. 

No system of interest, except the entire universe, can be 
thought of as an isolated system -- all subsets of the universe 
are in fact `open' systems, each surrounded by a larger system 
constituting its environment. The theory of open quantum 
systems provides a natural route for reconciliation of 
dissipation and decoherence with the process of quantization. 
In this picture, friction or damping comes about by the 
transfer of energy from the `small' system (the system of 
interest) to the `large' environment. The energy, once 
transferred, disappears into the environment and is not given 
back within any time of physical relevance. Ford, Kac and Mazur 
\cite{fkm65} suggested the first microscopic model describing 
dissipative effects in which the system was assumed to be 
coupled to a reservoir of an infinite number of harmonic 
oscillators. Interest in quantum dissipation, using the 
system-environment approach, was intensified by the works of 
Caldeira and Leggett \cite{cl83}, and Zurek \cite{wz91} among 
others. The path-integral approach, developed by Feynman and 
Vernon \cite{fv63}, was used by Caldeira and Leggett 
\cite{cl83}, and the reduced dynamics of the system of interest 
was followed taking into account the influence of its 
environment, quantified by the influence functional. In the 
model of the fluctuating or ``Brownian" motion of a quantum 
particle studied by Caldeira and Leggett \cite{cl83}, the 
coordinate of the particle was coupled linearly to the harmonic 
oscillator reservoir, and it was also assumed that the system 
and the environment were initially factorized. The treatment of 
the quantum Brownian motion has since been generalized to the 
physically reasonable initial condition of a mixed state of the 
system and its environment by Hakim and Ambegaokar \cite{ha85}, 
Smith and Caldeira \cite{sc87}, Grabert, Schramm and Ingold 
\cite{gsi88}, and by us for the case of a system in a 
Stern-Gerlach potential \cite{sb00}, and also for the quantum 
Brownian motion with nonlinear system-environment couplings 
\cite{sb03-2}. 

An open system Hamiltonian is of the QND type if the 
Hamiltonian $H_S$ of the system commutes with the Hamiltonian 
$H_{SR}$ describing the system-reservoir interaction, i.e., 
$H_{SR}$ is a constant of motion generated by $H_S$. 
Interestingly, such a system may still undergo decoherence or 
dephasing without any dissipation of energy \cite{gkd01,sgc96}. 

In this Letter, we study such QND `open system' Hamiltonians of 
particular interest in the context of decoherence in quantum 
computing, and obtain the propagators of the composite systems 
explicitly using path integral methods. The aim is to shed some 
light on the problem of QND measurement schemes. Can one draw 
upon any familiar symmetries to connect with the time-evolution 
operation of these QND systems? 

We take our system to be a two-level atom, or equivalently, a 
spin-1/2 system. We consider two types of environment, 
describable as baths of either oscillators or spins. One cannot 
in general map a spin-bath to an oscillator-bath (or vice 
versa); they constitute distinct 'universality classes' of 
quantum environment \cite{rpp00}. The first case of  
oscillator-bath models (originated by Feynman and Vernon 
\cite{fv63}) describes delocalized environmental modes. For the 
spin-bath, on the other hand, the finite Hilbert space of each 
spin makes it appropriate for describing the low-energy 
dynamics of a set of localized environmental modes. A 
difficulty associated with handling path integrals for spins 
comes from the discrete matrix nature of the spin-Hamiltonians. 
This difficulty is overcome by bosonizing the Hamiltonian by 
representing the spin angular momentum operators in terms of 
boson operators following Schwinger's theory of angular 
momentum \cite{schwin}. 

We then use the Bargmann representation \cite{vb47} for all the 
boson operators. The Schr\"{o}dinger representation of quantum 
states diagonalizes the position operator, expressing pure 
states as wave functions, whereas the Bargmann representation 
diagonalizes the creation operator $b^{\dagger}$, and expresses 
each state vector $|\psi \rangle$ in the Hilbert state ${\cal 
H}$ as an entire analytic function $f(\alpha)$ of a complex 
variable $\alpha$. The association $|\psi \rangle 
\longrightarrow f(\alpha)$ can be written conveniently in terms 
of the normalized coherent states $|\alpha \rangle$ which are 
the right eigenstates of the annihilation operator $b$: 
\begin{eqnarray}
b|\alpha \rangle & = & \alpha |\alpha \rangle , \nonumber \\ 
\langle \alpha '|\alpha \rangle & = & \exp \left( -\frac{1}{2}| 
\alpha '|^2 - \frac{1}{2}|\alpha |^2 + \alpha '^* \alpha 
\right) , \nonumber 
\end{eqnarray}
giving
\[ f(\alpha ) = e^{-|\alpha |^2/2} ~\langle \alpha ^* |\psi 
\rangle . \] 
We obtain the explicit propagators for these many-body systems 
from those of the expanded bosonized forms by appropriate 
projection. 

The propagators for the QND Hamiltonians with an oscillator 
bath and a spin bath are shown to be connected to the squeezing 
and rotation operators, respectively, which are both phase 
space area-preserving canonical transformations. This brings 
out an interesting analogy between the energy-preserving QND 
Hamiltonians and the homogeneous linear canonical 
transformations, which would need further systematic probing. 

We first take the case where the system is a two-level atom 
interacting with a bosonic bath of harmonic oscillators with a 
QND type of coupling. Such a model has been studied 
\cite{unruh95, ps96, dd95} in the context of the influence of 
decoherence in quantum computation. The total system evolves 
under the Hamiltonian, 
\begin{eqnarray}
H & = & H_S + H_R + H_{SR} \nonumber\\ & = & {\hbar \omega 
\over 2} \sigma_z + \sum\limits^M_{k=1} \hbar \omega_k 
b^{\dagger}_k b_k + \left( {\hbar \omega \over 2} \right) 
\sum\limits^M_{k=1} g_k (b_k + b^{\dagger}_k) \sigma_z.   
\label{h1} 
\end{eqnarray}
Here $H_S, H_R$ and $H_{SR}$ stand for the Hamiltonians of the 
system, reservoir, and system-reservoir interaction, 
respectively. We have made use of the equivalence of a 
two-level atom and a spin-1/2 system, $\sigma_x, \sigma_z$ 
denote the standard Pauli spin matrices and are related to the 
spin-flipping (or atomic raising and lowering) operators $S_+$ 
and $S_-$: $\sigma_x = S_+ + S_-$, $\sigma_z = 2 S_+ S_- - 1$. 
In (\ref{h1}) $b^{\dagger}_k, b_k$ denote the Bose creation and 
annihilation operators for the $M$ oscillators of frequency 
$\omega_k$ representing the reservoir, $g_k$ stands for the 
coupling constant (assumed real) for the interaction of the 
field with the spin. Since $[H_S, H_{SR}]=0$, the Hamiltonian 
(\ref{h1}) is of QND type. 

The explicit propagator $\exp (-{i H t \over \hbar})$ for the 
Hamiltonian (\ref{h1}) is obtained by using functional 
integration and bosonization \cite{gp86,sb03-1}, the details of 
which can be found elsewhere \cite{sb05-2}. In order to express 
the spin angular momentum operators in terms of boson 
operators, we employ Schwinger's theory of angular momentum 
\cite{schwin} by which any angular momentum can be represented 
in terms of a pair of boson operators with the usual 
commutation rules. The spin operators $\sigma_z$ and $\sigma_x$ 
can be written in terms of the boson operators $a_{\beta}$, 
$a_{\beta}^{\dagger}$ and $a_{\gamma}$, $a_{\gamma}^{\dagger}$ 
as 
\begin{eqnarray}
\sigma_z & = & a_{\gamma}^{\dagger} a_{\gamma} -  
a_{\beta}^{\dagger} a_{\beta}, \nonumber \\ \sigma_x & = & 
a_{\gamma}^{\dagger} a_{\beta} +  a_{\beta}^{\dagger} 
a_{\gamma} . \nonumber 
\end{eqnarray}
In the Bargmann representation \cite{vb47} the actions of $b$ 
and $b^{\dagger}$ are 
\begin{eqnarray}
b^{\dagger} f(\alpha) & = & \alpha^* f(\alpha), \nonumber\\ b 
f(\alpha) & = & {df(\alpha) \over d\alpha^*},   
\end{eqnarray}
where $|\alpha \rangle$ is the normalized coherent state. The 
spin operator becomes
\begin{equation}
\sigma_z \longrightarrow \left(  \gamma^* {\partial \over 
\partial \gamma^*} - \beta^* {\partial \over \partial \beta^*} 
\right).   
\end{equation}
Here the variable $\beta^*$ is associated with the spin-down 
state and the variable $\gamma^*$ with the spin-up state. 

The propagator for the bosonized form of (\ref{h1}) is given as 
a path integral over coherent state variables, and the required 
propagator is then obtained by appropriate projection, the 
amplitudes of which are given in a matrix form as \cite{sb05-2} 
\begin{eqnarray}
u_1 (\mbox{\boldmath $\alpha^*$}, \beta^*, \gamma^*, t; 
\mbox{\boldmath $\alpha'$}, \beta', \gamma', 0) & = & \exp 
\left\{ \sum\limits^M_{k=1} \alpha^*_k \alpha'_k e^{-
i\omega_kt} \right\} \nonumber\\ & & \times e^A \pmatrix{e^B & 
0 \cr 0 & e^{-B}} . \label{h2} 
\end{eqnarray}    
In the above equation {\boldmath $\alpha$} is a vector with 
components $\{\alpha_k\}$ and 
\begin{equation}
A = i \left( {\omega \over 2} \right)^2 \sum\limits^M_{k=1} 
{g^2_k \over \omega_k} t - \left( {\omega \over 2} \right)^2 
\sum\limits^M_{k=1} {g^2_k \over \omega^2_k} (1-e^{-
i\omega_kt}), \label{h3} 
\end{equation}
\begin{equation}
B = \sum\limits^M_{k=1} \phi_k \left( \alpha^*_k + \alpha'_k 
\right) + i {\omega \over 2} t, \label{h4} 
\end{equation}
\begin{equation}
\phi_k = {\omega \over 2} {g_k \over \omega_k} \left( 1-e^{-
i\omega_kt} \right). \label{h5} 
\end{equation} 
Here we associate the values $\alpha^*$ with time $t$ and 
$\alpha'$ with time $t=0$. The simple form of the last term on 
the right-hand side of (\ref{h2}) reveals the QND nature of the 
system-reservoir coupling. Since we are considering the unitary 
dynamics of the complete Hamiltonian (\ref{h1}) there is no 
decoherence, and the propagator (\ref{h2}) does not have any 
off-diagonal terms. In a treatment of the system alone, i.e., 
an open system analysis of Eq. (\ref{h1}) after the tracing 
over the reservoir degrees of freedom, it has been shown 
\cite{ps96} that the population, i.e., the diagonal elements of 
the reduced density matrix of the system remain constant in 
time while the off-diagonal elements that are a signature of 
the quantum coherences decay due to decoherence, as expected. 

Next we consider a variant of the Hamiltonian (\ref{h1}) 
wherein we include an external mode in resonance with the 
atomic transition, such as  
\begin{eqnarray}
H & = & {\hbar \omega \over 2} \sigma_z + \hbar \Omega 
a^{\dagger}a - {\hbar \Omega \over 2} \sigma_z \nonumber\\ & & 
+ \sum\limits^M_{k=1} \hbar \omega_k b^{\dagger}_k b_k + \left( 
{\hbar \omega \over 2} \right) \sum\limits^M_{k=1} g_k (b_k + 
b^{\dagger}_k) \sigma_z. \label{h7} 
\end{eqnarray}
Here
\begin{equation}
\Omega = 2 \vec{\epsilon}.\vec{d}^* ,
\end{equation}
where $\vec{d}$ is the dipole transition matrix element and 
$\vec{\epsilon}$ comes from the field strength of the external 
driving mode $\vec{E}_L(t)$ such that 
\begin{equation}
\vec{E}_L(t) = \vec{\epsilon} e^{-i\omega t} + \vec{\epsilon}~^* 
e^{i\omega t}. 
\end{equation}
The use of $\sigma_z$ instead of the usual $\sigma_x$ in the 
third term on the right-hand side of (\ref{h7}) allows us to 
have a QND Hamiltonian. Proceeding as before and introducing 
the symbol $\nu^*$ for the external mode $a^{\dagger}$ we 
obtain the amplitudes of the propagator for (\ref{h7}) as 
\begin{eqnarray}
u_2 (\nu^*, \mbox{\boldmath $\alpha^*$}, \beta^*, \gamma^*, t; 
\nu', \mbox{\boldmath $\alpha'$}, \beta', \gamma', 0) & = & 
\exp \left\{ \sum\limits^M_{k=1} \alpha^*_k \alpha'_k e^{-
i\omega_k t} \right\} \nonumber \\ & & \times \exp \left\{ 
\nu^* \nu' e^{-i\Omega t} \right\} ~ e^A \pmatrix{e^{B_2} & 0 
\cr 0 & e^{-B_2}} , \label{h8} 
\end{eqnarray}
where $A$ is as in Eq. (\ref{h3}),
\begin{equation}
B_2 = \sum\limits^M_{k=1} \phi_k (\alpha^*_k + \alpha'_k) + i 
\left( {\omega - \Omega \over 2} \right) t, \label{h9} 
\end{equation}
$\phi_k$ is as in Eq. (\ref{h5}). 

Now we consider the case where the reservoir is composed of 
spins or two-level systems, as has been dealt with by Shao and 
collaborators in the context of QND systems \cite{sgc96} and 
also quantum computation \cite{sh97}, and for a nanomagnet 
coupled to nuclear and paramagnetic spins \cite{rpp00}. The 
total Hamiltonian is taken as 
\begin{eqnarray}
H & = & H_S + H_R + H_{SR} \nonumber\\ & = & {\hbar \omega 
\over 2} S_z + \sum\limits^M_{k=1} \hbar \omega_k \sigma_{zk} + 
{\hbar \omega \over 2} \sum\limits^M_{k=1} c_k \sigma_{xk}S_z . 
\label{h10} 
\end{eqnarray}
Here we use $S_z$ for the system and $\sigma_{zk}, \sigma_{xk}$ 
for the bath. Since $[H_S, H_{SR}]=0$, we have a QND 
Hamiltonian. In the Bargmann representation, we associate the 
variable $\beta^*$ with the spin-down state and the variable 
$\gamma^*$ with the spin-up state for the bath variables, and 
we have 
\begin{eqnarray}
\sigma_z & \longrightarrow & \gamma^* {\partial \over \partial 
\gamma^*} - \beta^* {\partial \over \partial \beta^*}, 
\nonumber \\ \sigma_x & \longrightarrow & \gamma^* {\partial 
\over \partial \beta^*} + \beta^* {\partial \over \partial 
\gamma^*}. 
\end{eqnarray}
Similarly, the bosonization of the system variable gives 
\begin{equation}
S_z \longrightarrow \xi^* {\partial \over \partial \xi^*} - 
\theta^* {\partial \over \partial \theta^*} , 
\end{equation}
where the variable $\theta^*$ is associated with the spin-down 
state and the variable $\xi^*$ with the spin-up state. The 
amplitudes of the propagator for (\ref{h10}), in the Hilbert 
space of $H_R$, are obtained as 
\begin{eqnarray}
u_3 (\theta^*, \xi^*, \mbox{\boldmath $\beta^*$},  
\mbox{\boldmath $\gamma^*$}, t; \theta', \xi', \mbox{\boldmath 
$\beta'$}, \mbox{\boldmath $\gamma'$}, 0) & = & \prod^M_{k=1} 
\sum\limits^{\infty}_{n=0} (i\omega_k)^n \int\limits^t_0 
d\tau_n \int\limits^{\tau_{n}}_0 d\tau_{n-1}... 
\int\limits^{\tau_{2}}_0 d\tau_1 \nonumber \\ & & \times 
e^{i{\omega \over 2} S_z t} \pmatrix{\cos \Theta^{k(n)} & i 
\sin \Theta^{k(n)} \cr (-1)^ni \sin \Theta^{k(n)} & (-1)^n \cos 
\Theta^{k(n)}}. \label{h11} 
\end{eqnarray}
Here {\boldmath $\beta^*$}, {\boldmath $\gamma^*$} are vectors 
with components $\{ \beta_k\}$ and $\{ \gamma_k\}$, 
respectively, and 
\begin{equation}
\Theta^{k(n)} = {\omega \over 2} S_z c_k A_n,  \label{h12}
\end{equation}
\begin{equation}
A_n = \sum\limits^n_{j=1} (-1)^{j+1} 2\tau_j + (-1)^nt.  
\label{h13} 
\end{equation}
In Eq. (\ref{h11}), expanding the terms containing $S_z$ in the 
system space, we can see that out of the 16 amplitudes of the 
propagator for each mode of the reservoir, only the 
energy-conserving diagonal terms survive in Eq. (\ref{h11}) due 
to the QND nature of the system-reservoir coupling. 

We look closely at the forms of the propagators (\ref{h2}) and 
(\ref{h11}) of the QND type Hamiltonians (\ref{h1}) and 
(\ref{h10}), respectively. In the first case with an oscillator 
bath, Eq. (\ref{h2}) involves the matrix 
\[ \pmatrix{e^B & 0 \cr 0 & e^{-B}}, \]
where $B$ is given by Eq. (\ref{h4}). This can be used to 
generate the following transformation in phase space: 
\begin{equation}
\pmatrix{X \cr P} = \pmatrix{e^B & 0 \cr 0 & e^{-B}} \pmatrix{x 
\cr p}. \label{s1} 
\end{equation}
It can be easily seen from Eq. (\ref{s1}) that the Jacobian of 
the transformation is unity and it is a phase space 
area-preserving transformation. The first matrix on the 
right-hand side of (\ref{s1}) has the form of a `squeezing' 
operation \cite{km91}, which is an area-preserving (in phase 
space) canonical transformation coming out as an artifact of 
homogeneous linear canonical transformations \cite{bk05}. 

In the second case of a spin bath, Eq. (\ref{h11}) involves the 
matrix 
\begin{equation}
R \equiv \pmatrix{\cos \Theta^{k(n)} & i \sin \Theta^{k(n)} \cr 
(-1)^ni \sin \Theta^{k(n)} & (-1)^n \cos \Theta^{k(n)}}, 
\label{h16} 
\end{equation}
where $\Theta^{k(n)}$ is given by Eq. (\ref{h12}). For 
particular $n$ and $k$, we write $\Theta^{k(n)}$ as $\Theta$. 
For $n$ even, the above matrix (\ref{h16}) becomes 
\begin{equation}
\pmatrix{\cos \Theta & i\sin \Theta \cr i\sin \Theta & \cos 
\Theta} = e^{i\Theta \sigma_x}.  \label{h17} 
\end{equation}
Using the Campbell-Baker-Hausdorff identity \cite{qsrl} this 
matrix can be shown to transform the spin vector $\sigma = 
(\sigma_x, \sigma_y, \sigma_z)$ as 
\begin{equation}
e^{i\Theta \sigma_x} \pmatrix{\sigma_x \cr \sigma_y \cr 
\sigma_z} e^{-i\Theta \sigma_x} = \pmatrix{1 & 0 & 0 \cr 0 & 
\cos 2\Theta & -\sin 2\Theta \cr 0 & \sin 2\Theta & \cos 
2\Theta} \pmatrix{\sigma_x \cr \sigma_y \cr \sigma_z},     
\label{h18} 
\end{equation}
i.e., the abstract spin vector is `rotated' about the $x$-axis 
by an angle $2\Theta$. For $n$ odd, (\ref{h16}) becomes (again 
writing $\Theta^{k(n)}$ for particular $n$ and $k$ as $\Theta$) 
\begin{equation}
\pmatrix{\cos \Theta & i\sin \Theta \cr -i\sin \Theta & -\cos 
\Theta} = \sigma_z \pmatrix{\cos \Theta & i\sin \Theta \cr 
i\sin \Theta & \cos \Theta} = \sigma_z e^{i\Theta \sigma_x}.   
\label{h19} 
\end{equation}
Thus the $n$-odd matrix is related to the $n$-even matrix by 
the spin-flipping energy. The above matrix transforms the spin 
vector $\sigma$ as 
\begin{equation}
\sigma_z e^{i\Theta \sigma_x} \pmatrix{\sigma_x \cr \sigma_y 
\cr \sigma_y} e^{-i\Theta \sigma_x} \sigma_z = e^{i\pi} 
\pmatrix{1 & 0 & 0 \cr 0 & \cos 2\Theta & \sin 2\Theta \cr 0 & 
\sin 2\Theta & -\cos 2\Theta} \pmatrix{\sigma_x \cr \sigma_y 
\cr \sigma_z}.  \label{r1} 
\end{equation}
It can be easily seen from the right-hand side of the 
Eq. (\ref{r1}) that the determinant of the transformation of 
the spin vectors brought about by the $n$-odd matrix 
(\ref{h19}) has the value unity. It is well known that the 
determinant of a rotation matrix is unity \cite{gms}. Thus we 
see that the above transformation has the form of a rotation. 
Specifically, it can be seen that 
\begin{equation}
\sigma_z \pmatrix{\cos 2\Theta & \sin 2\Theta \cr \sin 2\Theta 
& -\cos 2\Theta} = \pmatrix{ \cos 2\Theta & \sin 2\Theta \cr -
\sin 2\Theta & \cos 2\Theta}, 
\end{equation} 
and
\begin{equation}
\pmatrix{\cos 2\Theta & \sin 2\Theta \cr -\sin 2\Theta & \cos 
2\Theta}^T = \pmatrix{ \cos 2\Theta & -\sin 2\Theta \cr \sin 
2\Theta & \cos 2\Theta}.   \label{h20} 
\end{equation}
Here $T$ stands for the transpose operation. From the above it 
is seen that the matrix (\ref{h16}) has the form of the 
operation of `rotation', which is also a phase space 
area-preserving canonical transformation \cite{km91} and comes 
out as an artifact of homogeneous linear canonical 
transformations \cite{bk05}. Any element of the group of 
homogeneous linear canonical transformations can be written as 
a product of a unitary and a positive transformation 
\cite{barg, bk052}, which in turn can be shown to have unitary 
representations (in the Fock space) of rotation and squeezing 
operations, respectively \cite{bk05}. It is interesting that 
the propagators for the Hamiltonians given by Eqs. (\ref{h1}) 
and (\ref{h10}), one involving a two-level system coupled to a 
bath of harmonic oscillators and the other with a bath of 
two-level systems, are analogous to the squeezing and rotation 
operations, respectively. 

In conclusion, we have investigated the forms of the 
propagators of some QND Hamiltonians commonly used in the 
literature, for example, for the study of decoherence in 
quantum computers. We have calculated the propagators for the 
Hamiltonian of a two-level system interacting with a bosonic 
bath of harmonic oscillators (and its variant wherein we have 
included an external mode in resonance with the atomic 
transition), as also a spin bath of two-level systems. In each 
case the system-bath interaction is of QND type, i.e., the 
Hamiltonian of the system commutes with the Hamiltonian 
describing the system-bath interaction. We have found an 
interesting analogue of the propagators of these many-body 
Hamiltonians to squeezing and to rotation, respectively. Every 
homogeneous linear canonical transformation can be factored 
into the rotation and squeezing operations and these cannot in 
general be mapped from one to the other -- just as one cannot 
in general map a spin bath to an oscillator bath (or vice 
versa) -- but together they span the class of homogeneous 
linear canonical transformations and are `universal'. Squeezing 
and rotation, being artifacts of homogeneous linear canonical 
transformations, are both phase-space area-preserving 
transformations, and thus this implies a curious analogy 
between the energy-preserving QND Hamiltonians and the 
homogeneous linear canonical transformations.  This insight 
into the structure of the QND systems would hopefully lead to 
future studies into this domain. 

It is a pleasure to acknowledge useful discussions with Joachim 
Kupsch. The School of Physical Sciences, Jawaharlal Nehru 
University, is supported by the University Grants Commission, 
India, under a Departmental Research Support scheme.

\end{document}